\documentstyle[12pt,epsfig]{article}        % Simple LaTeX, use with DINA4 as follows
\newlength{\dinwidth}
\newlength{\dinmargin}
\def\lsim{\mathrel{\rlap{\lower4pt\hbox{\hskip1pt$\sim$}}\raise1pt\hbox{$<$}}}
\def\gsim{\mathrel{\rlap{\lower4pt\hbox{\hskip1pt$\sim$}}
    \raise1pt\hbox{$>$}}}                % greater than or approx. symbol
\newcommand{\bq}{\begin{equation}}
\newcommand{\eq}{\end{equation}}

\begin{document}
% Begin of extra titlepage for the preprint version
\begin{titlepage}

\large
\normalsize
\begin{flushleft}
{\tt hep-ph/9609nnn}\\
September 1996
\end{flushleft}
\vspace*{4cm}
\begin{center}
\LARGE
{\bf QCD corrections to $F_L(x,Q^2)$} \\

\vspace{3cm}
\large
Johannes Bl\"umlein and Stephan Riemersma
\\
\vspace{2.5cm}
\large {\it
 DESY--Zeuthen \\

\vspace{0.1cm}
Platanenallee 6, D--15735 Zeuthen, Germany }\\
\vspace{\fill}
\normalsize
{\bf Abstract} \\
\end{center}
\noindent
We perform a
 numerical study  of the QCD corrections to the structure
function $F_L(x,Q^2)$ in the HERA energy range.
The $K$--factors are of $O(30 \%)$ and larger in parts of the kinematic
range. The relative
corrections to $F_L^{c\overline{c}}$ turn out to be
scale dependent and partially
compensate contributions to the massless
terms.
\vfill
\noindent
\normalsize
\begin{center}
{
\sf
Contribution to the Proceedings of the 1996 HERA Physics
Workshop}
\end{center}

\end{titlepage}
% End of extra titlepage for the preprint version
\vspace*{1cm}
\begin{center}
\begin{Large}
\begin{bf}
QCD corrections to $F_L(x,Q^2)$ \\
\end{bf}
\end{Large}
\vspace*{5mm}
\begin{large}

Johannes Bl\"umlein and Stephan Riemersma
\end{large}

DESY--Zeuthen,  Platanenallee~6,~D-15735~Zeuthen, Germany\\
\end{center}
\begin{quotation}
\noindent
{\bf Abstract:}~We perform a
 numerical study  of the QCD corrections to the structure
function $F_L(x,Q^2)$ in the HERA energy range.
The $K$--factors are of $O(30 \%)$ and larger in parts of the kinematic
range. The relative
corrections to $F_L^{c\overline{c}}$ turn out to be
scale dependent and partially
compensate contributions to the massless
terms.
\end{quotation}

%%%%%%%%%%%%%%%%%%%%%%%%%%%%%%%%%%%%%%%%%%%%%%%%%%%%%%%%%%%%%%%%%%%%%%%%%
%\section{Introduction}
%\label{sect1}
%%%%%%%%%%%%%%%%%%%%%%%%%%%%%%%%%%%%%%%%%%%%%%%%%%%%%%%%%%%%%%%%%%%%%%%%%
\noindent
The longitudinal structure function in deep inelastic scattering,
$F_L(x,Q^2)$, is one of the observables from
which the
gluon density can be unfolded. 
In leading order (LO)~\cite{R1} it is given by
%------------------------------------------------------------------------
\begin{equation}
\label{fl1}
F_L^{ep}(x,Q^2) = \frac{\alpha_s(Q^2)}{\pi}
\left \{ \frac{4}{3} c_{L,1}^q(x) \otimes F_2^{ep}(x,Q^2) + 2 \sum_q
c_{L,1}^g(x,Q^2) \otimes [
xG(x,Q^2)] \right \}
\end{equation}
%------------------------------------------------------------------------
with
%------------------------------------------------------------------------
\begin{equation}
\label{fl2}
c_{L,1}^q(x) = x^2~~~~~~~c_{L,1}^g(x) = x^2 (1-x),
\end{equation}
%------------------------------------------------------------------------
and $\otimes$ denoting the Mellin convolution.
Eq.~(\ref{fl1}) applies for light quark flavours. Due to the power
behaviour of the coefficient functions $c_{L,1}^{q,g}(x)$,
an approximate relation for the gluon density at small $x$
%------------------------------------------------------------------------
\begin{equation}
\label{fl3}
xG(x, Q^2) \simeq \frac{3}{5} \times 5.85 \left \{ \frac{3\pi}
{4 \alpha_s(Q^2)} F_L(0.4x, Q^2) - \frac{1}{2} F_2(0.8x, Q^2) \right \},
\end{equation}
%------------------------------------------------------------------------
has been used to derive a simple estimate for $xG(x,Q^2)$ in the
past~\cite{R2}.
Heavy quark contributions and the next-to-leading order (NLO)
QCD corrections
complicate the unfolding of the gluon density using $F_L(x,Q^2)$
and have to be
 accounted for in terms of $K$-factors.
In the present note, these contributions are studied numerically for the
HERA energy range.

The NLO corrections for the case of light quark flavours were calculated
in ref.~\cite{R3} and the LO and NLO contributions for the heavy
flavour terms were derived in refs.~\cite{R4} and \cite{R5},
respectively. While in LO the heavy flavour part of $F_L(x,M^2)$
is only due to $\gamma^* g$ fusion, in NLO also light quark terms
contribute. Moreover, the choice  of the factorization scale $M^2$
happens to affect
$F_L^{Q\overline{Q}}(x,M^2)$ substantially.

\vspace{5mm}
\noindent
%%%%%%%%%%%%%%%%%%%%%%%%%%%%%%%%%%%%%%%%%%%%%%%%%%%%%%%%%%%%%%%%%%%%%%
{\large\bf Light flavour contributions }\\
%%%%%%%%%%%%%%%%%%%%%%%%%%%%%%%%%%%%%%%%%%%%%%%%%%%%%%%%%%%%%%%%%%%%%%

\vspace{1mm}
\noindent
The leading order contributions to $F_L(x,Q^2)$ are shown in Figure~1
for $x \geq 10^{-4}$ and $10 \leq Q^2 \leq 500~{\rm GeV}^2$.
Here and in the following we refer to the CTEQ
 parametrizations~\cite{R6} and assume $N_f = 4$. We also show the
 quarkonic contributions which are suppressed by one order of magnitude
 against the gluonic ones in the small $x$ range. The ratio of the
 NLO/LO contributions is depicted in figure~2. Under the above 
conditions, it exhibits a fixed point at $x \sim 0.03$. Below, the
correction grows for rising $Q^2$ from $K = 0.9$ to 1 for $x = 10^{-4}$,
$Q^2 \epsilon [10, 500]~{\rm GeV^2}$. Above, its behaviour is reversed.
The correction factor $K$ rises for large values of $x$. For $x \sim
0.3$ it reaches e.g. $1.4$ for $Q^2 = 10~{\rm GeV}^2$.
In NLO the quarkonic contributions are suppressed similarly as in the
LO case at small $x$ and contribute to $F_L$ by $15 \%$ if only
light flavours are assumed.

\vspace{5mm}
\noindent
%%%%%%%%%%%%%%%%%%%%%%%%%%%%%%%%%%%%%%%%%%%%%%%%%%%%%%%%%%%%%%%%%%%%%%
{\large\bf Heavy flavour contributions }\\
%%%%%%%%%%%%%%%%%%%%%%%%%%%%%%%%%%%%%%%%%%%%%%%%%%%%%%%%%%%%%%%%%%%%%%

\vspace{1mm}
\noindent
The heavy flavour contributions to $F_L$ are shown in figures~3 and 4,
comparing the results for the choices of the factorization scale
$M^2 = 4 m_c^2$ and $M^2 = 4 m_c^2 +Q^2$, with $m_c = 1.5~{\rm GeV}$.
Here we used again parametrization~\cite{R6} for the description of
the parton densities but referred to three light flavours only unlike
the case in the previous section. The comparison of Figures~3a and 4a
shows that the NLO corrections are by far less sensitive to the
choice of the factorization scale than the LO results.
Correspondingly the $K_{c\overline{c}}-$factors
 $F_L^{c\overline{c}}(NLO)
 /F_L^{c\overline{c}}(LO)$ are strongly scale dependent. Note that the
ratios $K_{c\overline{c}}$ and $K$ behave different and compensate
each other
partially. Thus the overall correction depends on the heavy-to-light
flavour composition of $F_L(x,Q^2)$.

\vspace{5mm}

In summary we note that the NLO corrections to $F_L$ are large. Partial
compensation between different contributions can emerge. For 
an unfolding of the gluon density from $F_L(x,Q^2)$ the NLO corrections
are indispensable.

%%%%%%%%%%%%%%%%%%%%%%%%%%%%%%%%%%%%%%%%%%%%%%%%%%%%%%%%%%%%%%%%%%%%%%%%%

%%%%%%%%%%%%%%%%%%%%%%%%%%%%%%%%%%%%%%%%%%%%%%%%%%%%%%%%%%%%%%%%%%%%%%%%
%%%%%%%%%%%%%%%%%%%%%%%%%%%%%%%%%%%%%%%%%%%%%%%%%%%%%%%%%%%%%%%%%%%%%%%%
\newpage
\twocolumn
\noindent
\mbox{\epsfig{file=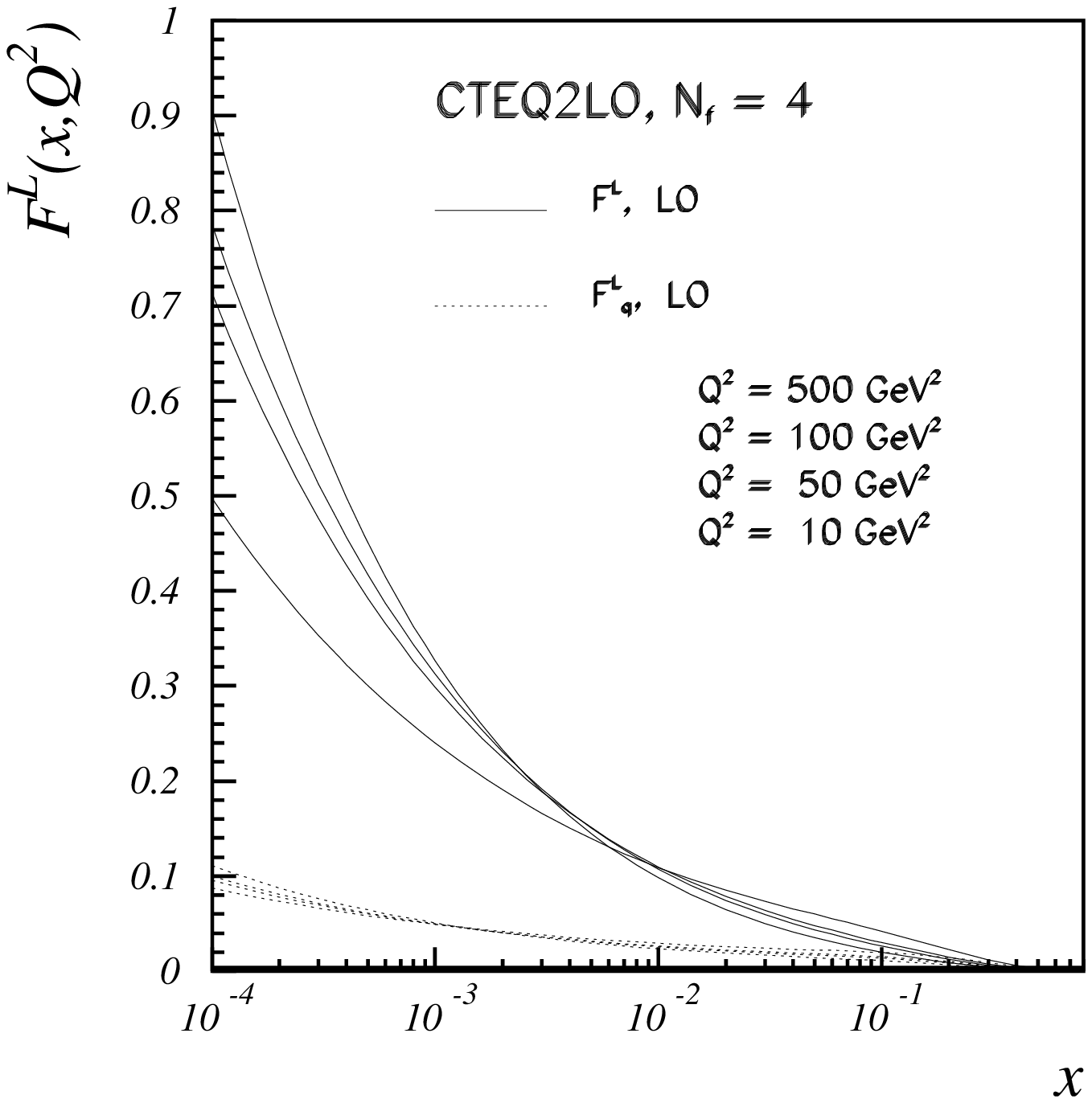,height=9cm,width=9cm}}
\small
\noindent
\sf Figure~1 : Leading order contributions to $F_L(x,Q^2)$. \\
\normalsize

\vspace{3mm}
\noindent
\mbox{\epsfig{file=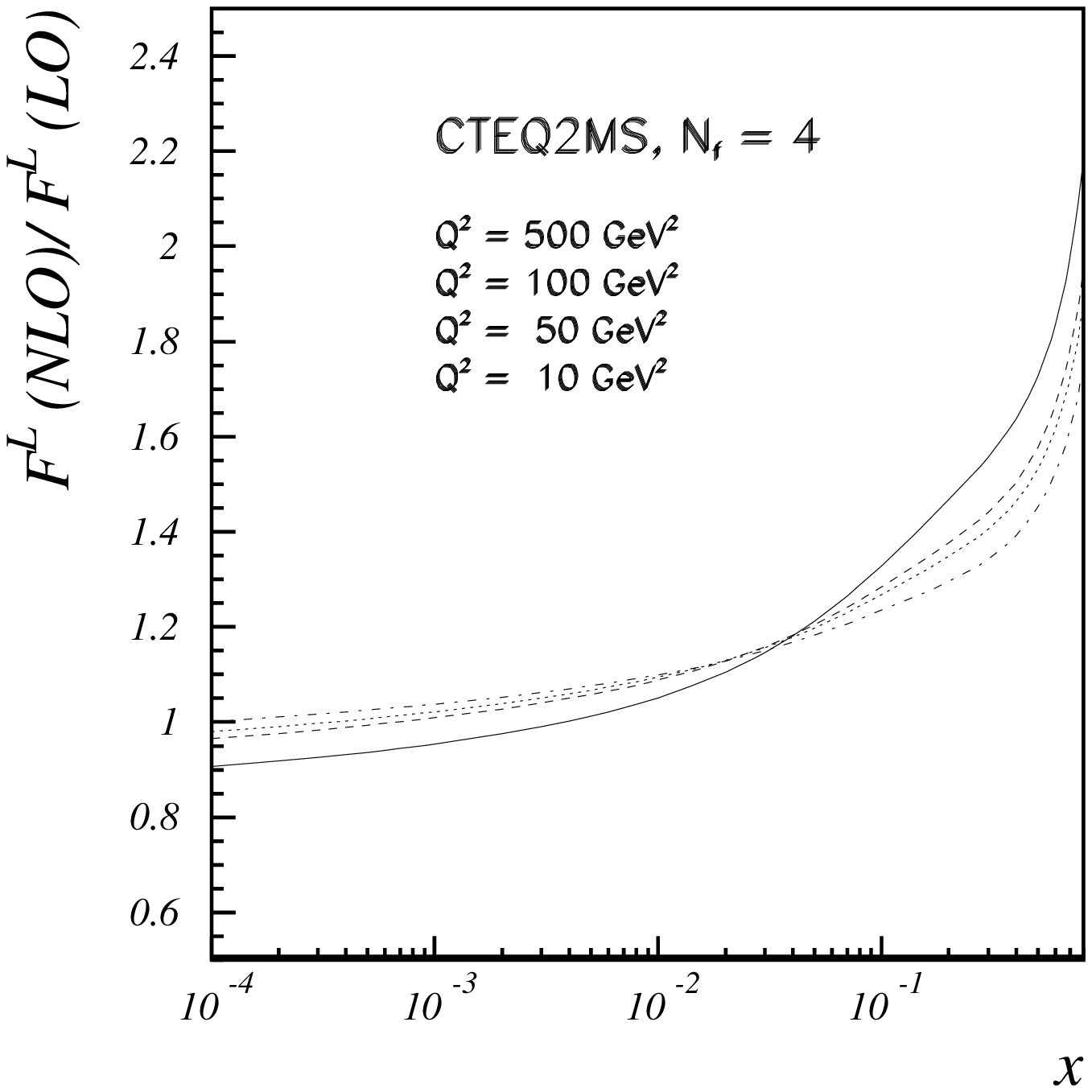,height=9cm,width=9cm}}
\small
\noindent
\sf Figure~2 : The NLO correction factor for $F_L(x,Q^2)$ in the
case of four light flavours.

\newpage
\mbox{\epsfig{file=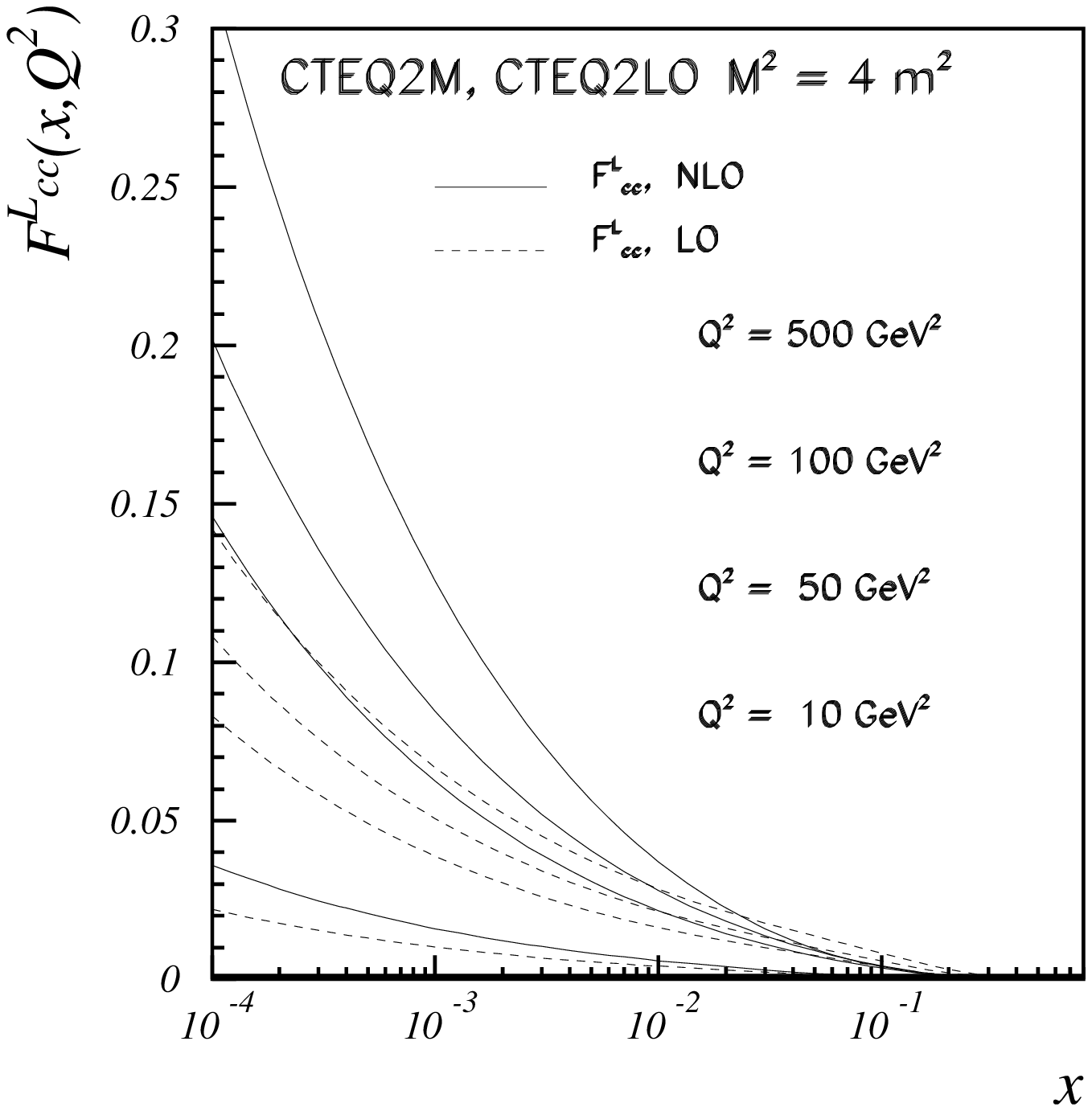,height=9cm,width=9cm}}
\noindent
\small
\sf Figure~3a : LO and NLO $c\overline{c}$ contributions to
$F_L(x,Q^2)$. The factorization scale is set to $M^2 = 4 m_c^2$.

\vspace{3mm}
\noindent
\mbox{\epsfig{file=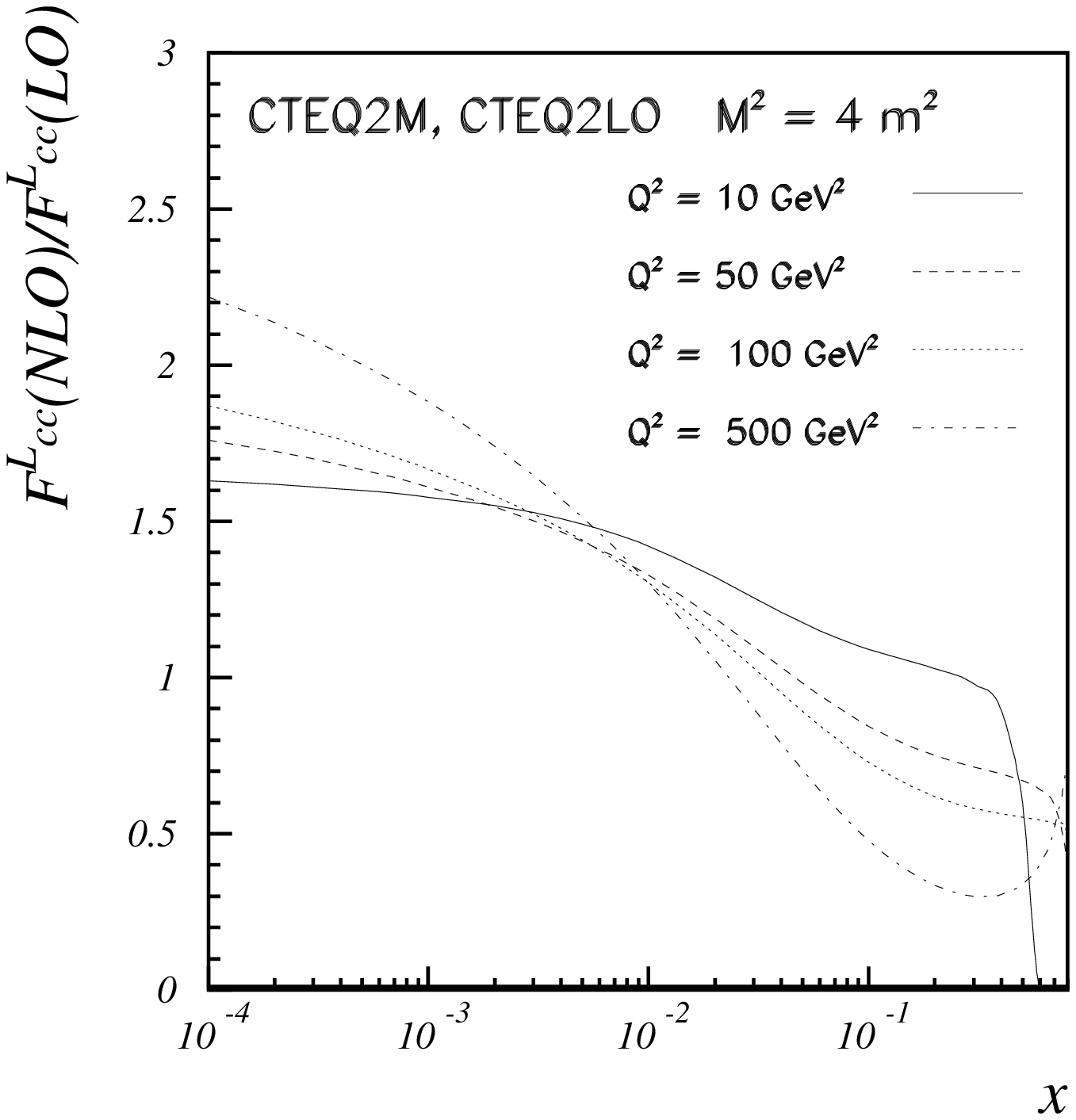,height=9cm,width=9cm}}
\noindent
\small
\sf Figure~3b : Ratio of the NLO to LO
$c\overline{c}$ contributions to
$F_L(x,Q^2)$. The factorization scale is set to $M^2 = 4 m_c^2$.

\vspace{4mm}
\noindent
\newpage
\noindent
\mbox{\epsfig{file=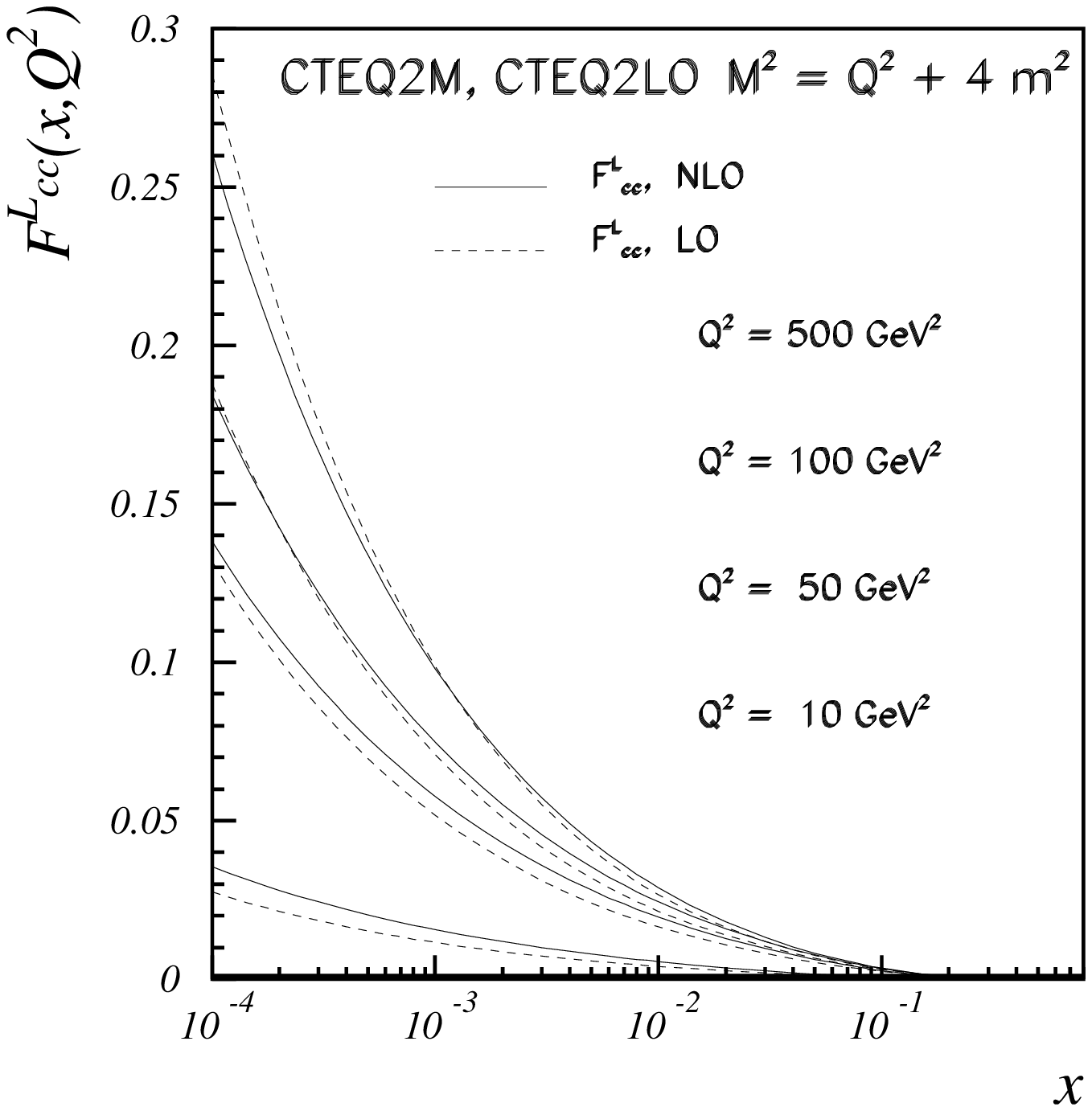,height=9cm,width=9cm}}
\noindent
\small
\sf Figure~4a : Same as in Fig.~3a but for choosing
the factorization scale   $M^2 = 4 m_c^2 + Q^2$.

\newpage
\noindent
\mbox{\epsfig{file=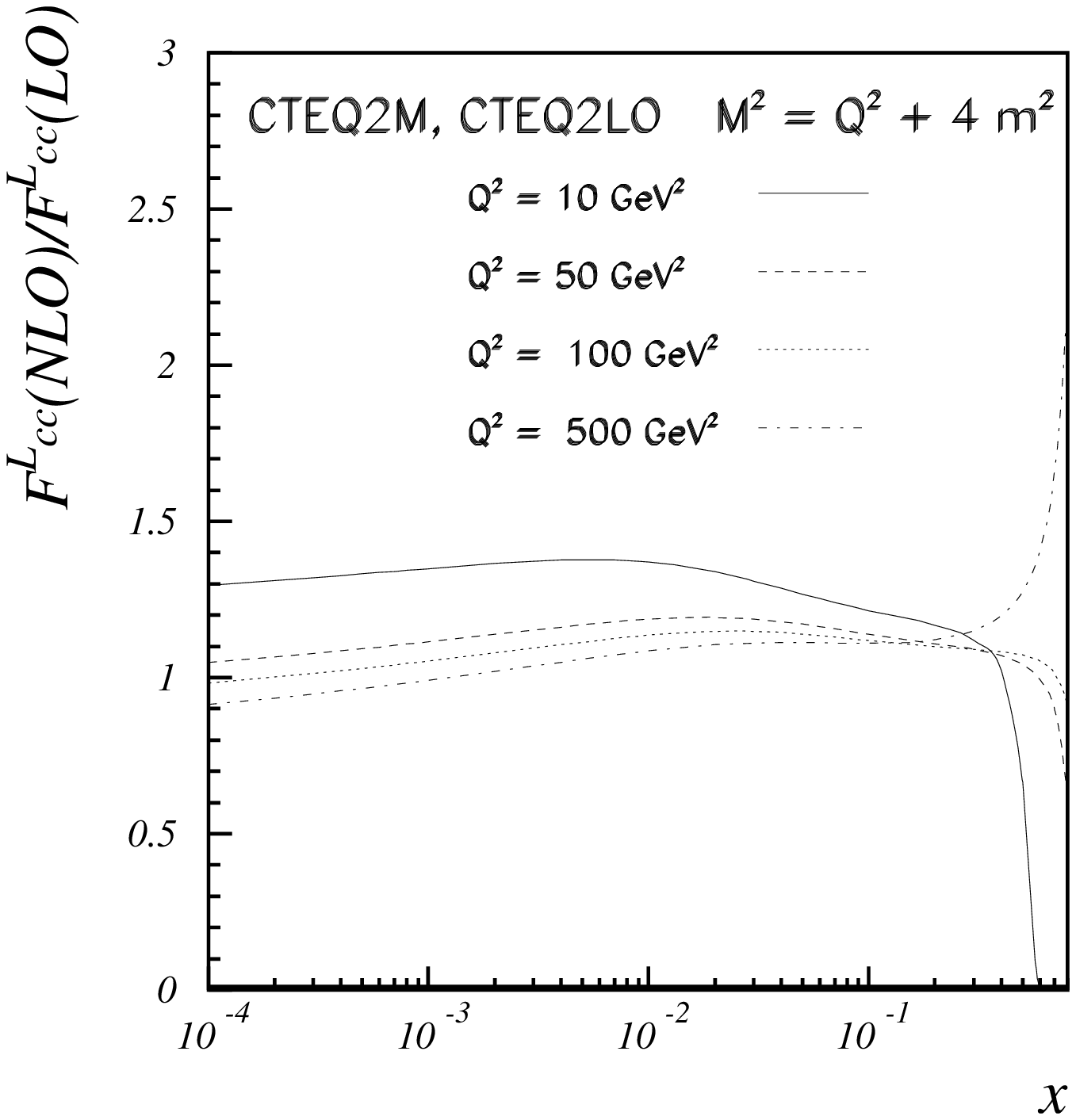,height=9cm,width=9cm}}
\noindent
\small
\sf Figure~4b : Same as in Fig.~3b but for choosing
the factorization scale   $M^2 = 4 m_c^2 + Q^2$.

\vspace{4mm}
\noindent
\end{document}